\documentclass[prb, aps, showpacs, showkeys, amsmath, amssymb, superscriptaddress, notitlepage, twocolumn]{revtex4-1}

\usepackage{graphicx}								
\usepackage[dvipsnames, usenames]{color}
\usepackage[english]{babel}

\usepackage{ifthen}
\newboolean{app}
\setboolean{app}{true}

\graphicspath{ {./Pictures/} }

\usepackage[normalem]{ulem} 	

\newcommand{\bra}[1]{\ensuremath{\left\langle #1\right|}}
\newcommand{\ket}[1]{\ensuremath{\left|#1\right\rangle}}

\newcommand{\mean}[1]{\ensuremath{\left\langle #1\right\rangle}}
\newcommand{\hc}{^{\dagger}}							
\newcommand{\op}[1]{\hat{#1}}							
\newcommand{\ee}{\mathrm{e}}						
\newcommand{\ii}{\mathrm{i}}							
\renewcommand{\H}[0]{\hat{H}}							
\newcommand{\nn}{\nonumber}							
\newcommand{\abs}[1]{\ensuremath{ \left| #1 \right| }}		
\newcommand{\abss}[1]{\ensuremath{ \left| #1 \right|^{2} }}	

\renewcommand{\l}[0]{\left}
\renewcommand{\r}[0]{\right}
\newcommand{\dvpart}[2]{\frac{\partial #1}{\partial #2}}		

\begin{document}

\title{
	Interacting two-level defects as sources of fluctuating high-frequency noise in superconducting circuits
	}

\author{Clemens M\"uller}
\affiliation{ARC Centre of Excellence for Engineered Quantum Systems, School of Mathematics and Physics, University of Queensland, Brisbane, Queensland 4072, Australia}
\affiliation{D\'epartement de Physique, Universit\'e de Sherbrooke, Sherbrooke, Qu\'ebec, Canada J1K 2R1}

\author{J\"urgen Lisenfeld}
\affiliation{Physikalisches Institut, Karlsruhe Institute of Technology, Karlsruhe, Germany}

\author{Alexander Shnirman}
\affiliation{Institut f\"ur Theorie der Kondensierten Materie, Karlsruhe Institute of Technology, Karlsruhe, Germany}
\affiliation{L. D. Landau Institute for Theoretical Physics RAS, 
Kosygina street 2, 119334 Moscow, Russia}

\author{Stefano Poletto}
\altaffiliation[present address: ]{QuTech Advanced Research Center and Kavli Institute of Nanosicence, Delft University of Technology, Lorentzweg 1, 2628 CJ Delft, The Netherlands}
\affiliation{IBM T.J. Watson Research Center, Yorktown Heights, New York 10598, USA}

\date{\today}

\begin{abstract}
	Since the very first experiments, superconducting circuits have suffered from strong coupling to environmental noise, destroying quantum coherence and degrading performance.
	In state-of-the-art experiments, it is found that the relaxation time of superconducting qubits fluctuates as a function of time. 
	We present measurements of such fluctuations in a 3D-transmon circuit and develop a qualitative model 
	based on interactions within a bath of background two-level systems (TLS) which emerge from defects in the device material.
	In our model, the time-dependent noise density acting on the qubit emerges from its near-resonant coupling to high-frequency TLS which experience energy fluctuations 
	due to their interaction with thermally fluctuating TLS at low frequencies. 
	We support the model by providing experimental evidence of such energy fluctuations observed in a single TLS in a phase qubit circuit. 
\end{abstract}


\maketitle

\section{Introduction}
Superconducting qubits~\cite{Clarke:N:2008} are well on the way towards achieving the prerequisites for fault-tolerant quantum computation schemes~\cite{Barends:N:2014, Chow:NC:2014, Saira:PRL:2014}.
With the advent of highly coherent superconducting circuits for quantum applications, previously neglected sources of environmental noise become important. 
One such cause of decoherence is spurious two-level systems (TLS), 
which are believed to be present in large numbers in the amorphous dielectric oxide layer covering the superconducting films~\cite{Simmonds:PRL:2004, Shalibo:PRL:2010}.
Ensembles of TLS naturally explain the low-temperature properties of glasses~\cite{Phillips:JLTP:1972, Anderson:PM:1972} 
and are used as a universal model for $1/f$-type low-frequency noise in electric circuits~\cite{Dutta:RMP:1981, Paladino:RMP:2013}.

Virtually all designs of superconducting qubits tested so far show a pronounced frequency dependence in their relaxation rates~\cite{Astafiev:PRL:2004, Ithier:PRB:2005, Barends:PRL:2013, Paik:PRL:2011}, 
which indicates strongly coloured high-frequency noise acting on the circuits~\cite{Shnirman:PRL:2005}.
A natural explanation of these observations relies on weak interactions between the circuit dynamics and spurious environmental TLS, possibly located in the disordered dielectric covering these circuits.
For coupling strengths that are much weaker than the individual decoherence rates of qubit and defect, the effect of the TLS on the qubit will be that of a strongly peaked high-frequency noise spectrum.
In other experiments, strongly coupled coherent TLS are often found to cause avoided level crossings in superconducting circuits 
which include Josephson junctions~\cite{Simmonds:PRL:2004, Shalibo:PRL:2010}.
Those TLS are believed to reside in the dielectric forming the tunnelling barrier inside the circuits Josephson junctions, enabling their stronger coupling to the circuit dynamics.
Otherwise they are conjectured to be of the same origin as the TLS observed as resonances in the high-frequency noise spectrum.
Using superconducting qubits as probes, it is possible to fully characterise the properties of the strongly coupled defects, 
for example by measuring their level-structure and coherence times~\cite{Bushev:PRB:2010, Lisenfeld:PRL:2010,Cole:APL:2010}. 

\begin{figure}[htbp]
	\begin{center}
		\includegraphics[width=.85\columnwidth]{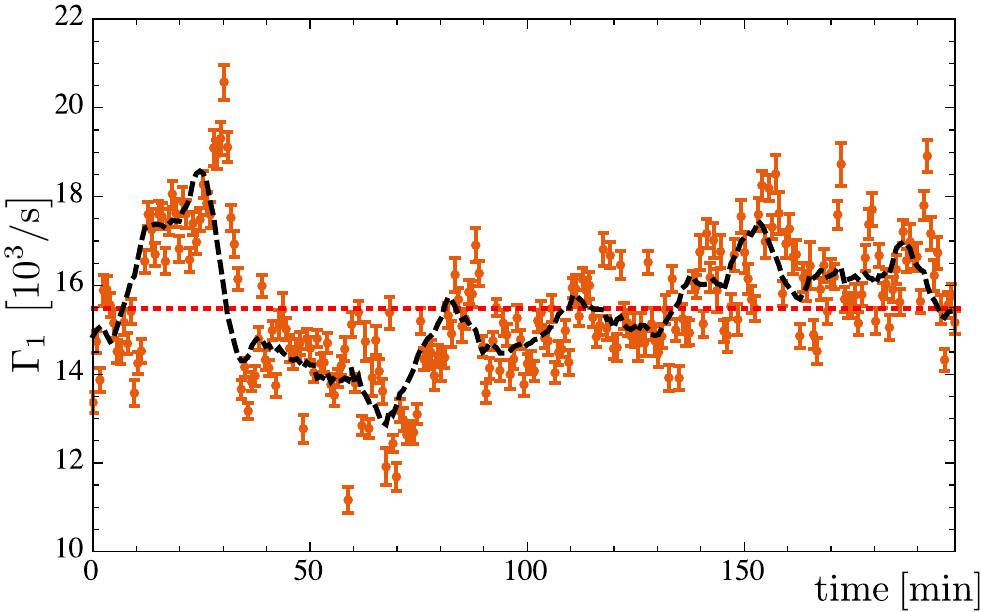}
		\caption{(Color online) Fluctuations in time of the relaxation rate $\Gamma_{1}$ of a superconducting 3D-transmon qubit.
			Errorbars show the 95\% confidence interval of the fits, the dotted red line indicates the mean value of all measurements and the dashed black line is a moving average over 10 samples
			to emphasize the multivalued character of the jumps. 
			Each individual point in this plot required a measurement time of $\sim 1$~min.
		}
		\label{fig:Gamma1Time}
	\end{center}
\end{figure}

In this work, we discuss the origin of time-dependent fluctuations in the energy relaxation time $T_{1}$, which are observed in superconducting  2D-transmons~\cite{Macha:2014}, 
flux qubits~\cite{Clarke:2014} and 3D-transmons~\cite{Paik:PRL:2011}, as shown in Fig.~\ref{fig:Gamma1Time}. 
The paper is organised as follows: Section~\ref{sec:Mot} starts by motivating this work and our theoretical model for time-dependent fluctuations in the relaxation rate of superconducting circuits.
Section~\ref{sec:Exp} then describes the experiments from which our data originate. In the following part, Sec.~\ref{sec:Theo}, we develop the model and present the main results. 
The discussion in Sec.~\ref{sec:Disc} presents implications and possible tests of the model and considers possible alternative explanations of the data. 
The paper is followed by appendices summarising details of the experiments, additional experimental data, and providing more details on the theoretical calculations.

\section{Motivation\label{sec:Mot}}
\begin{figure}[htbp]
	\begin{center}
		\includegraphics[width=.95\columnwidth]{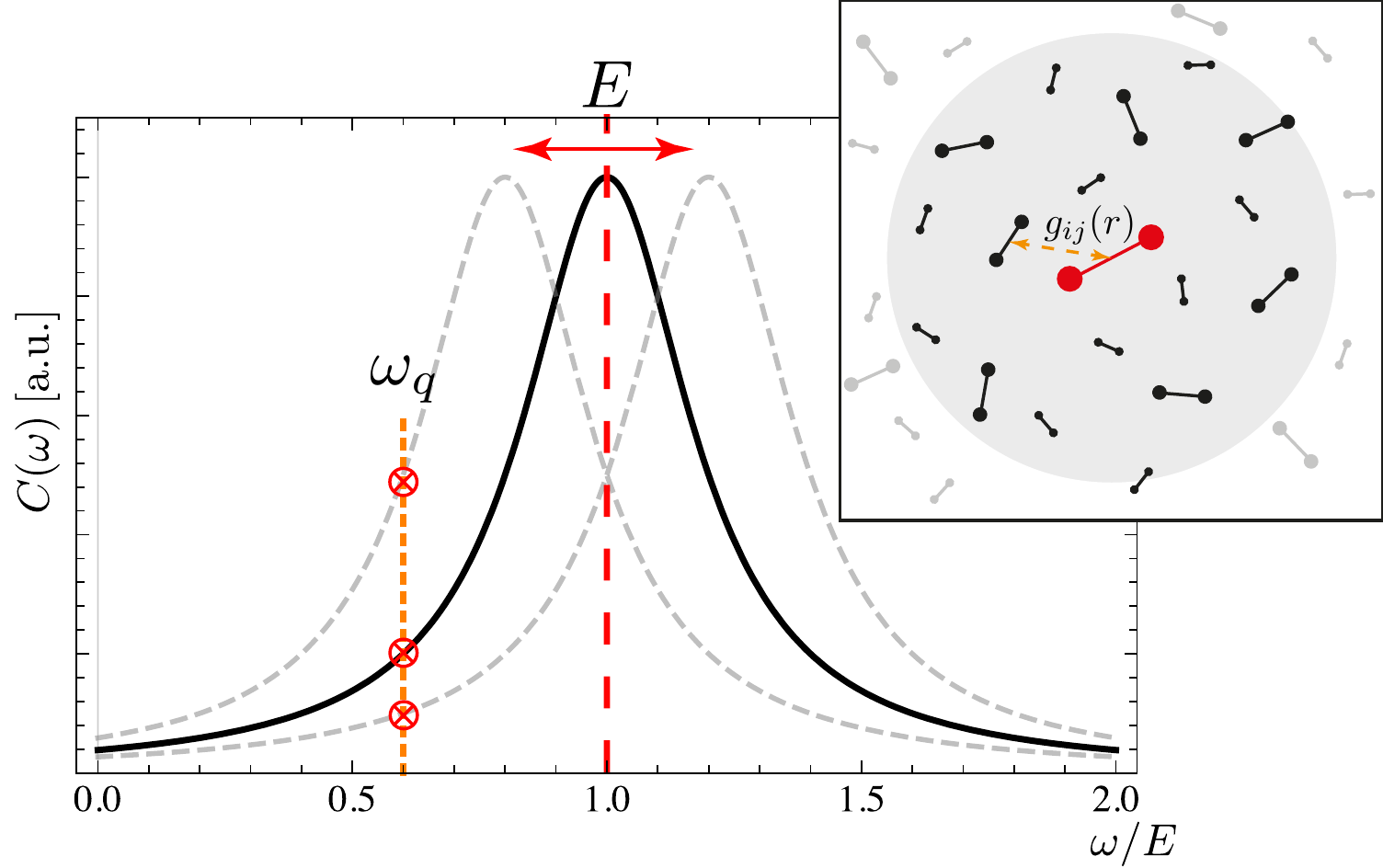}
		\caption{(Color online) Illustration of the conjectured mechanism behind fluctuations in $\Gamma_{1}$. 
			We plot the noise spectral density $C(\omega)$, Eq.~\eqref{eq:COmega}, of a single high-frequency TS as a function of frequency $\omega$. 
			The qubit level splitting is indicated as $\omega_{q}$ and the TS energy as $E$.
			Fluctuations in $E$, as indicated by the arrow and the dashed contours, 
			may cause strong changes in the noise spectral density at the qubit frequency, 
			leading to significant changes in the qubit relaxation rate $\Gamma_{1} \propto C(\omega_{q})$.
			The inset shows an illustration of the interaction between a central high-frequency TS (red, centre) with a surrounding bath of low-frequency TF (black),
			where the interaction is limited to a small spatial range indicated by the grey-shaded region.
		}
		\label{fig:TLSNoise}
	\end{center}
\end{figure}

\begin{figure}[htbp]
	\begin{center}
		\includegraphics[width=.85\columnwidth]{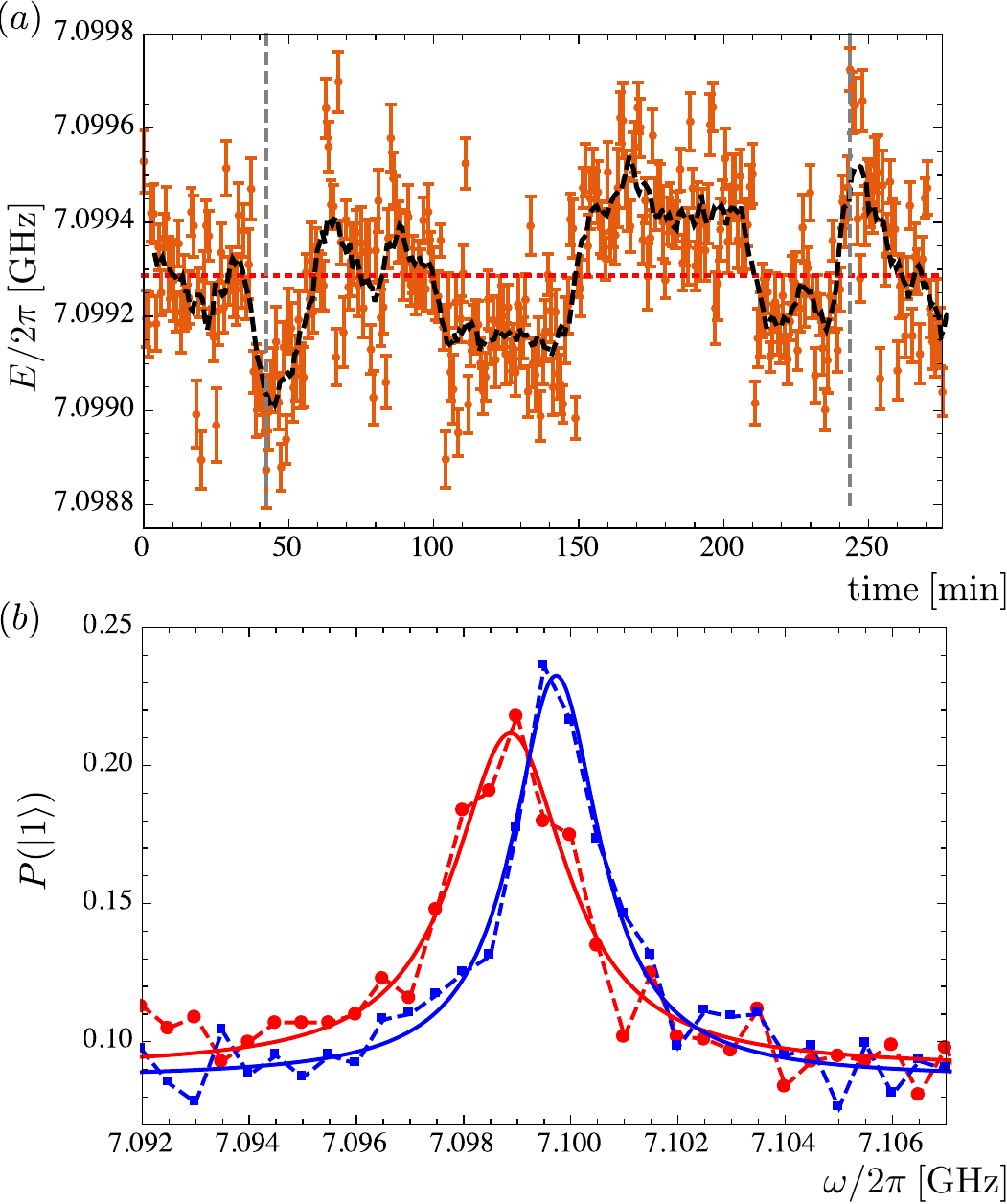}
		\caption{(Color online) (a) Change in TS energy $E$ as a function of time measured in a superconducting phase qubit.
			Errorbars indicate the 95\% confidence interval of the fits, the red dotted line is the average over the samples shown and the black dashed line is a moving average over 10 samples.
			Panel (b) shows two Lorentzians in the escape probability of the qubit $P(\ket1)$ at two different times as an example of the change in TS energy. 
			Here the dots are the raw data and the solid lines are the result of a fit to the data. Vertical dashed lines in (a) indicate the measurement times for the two curves shown.
		}
		\label{fig:TLSEnergy}
	\end{center}
\end{figure}

Qubit relaxation may occur through its weak coupling to environmental TLS whose characteristic eigenenergies are comparable to the qubit's energy splitting.
The environmental noise spectral density originating from coupling to a single such TLS is strongly peaked around its eigenfrequency. 
A natural approach to explain the fluctuations in the qubit relaxation rate is thus to assume random changes in the energy splitting of individual two-level defects; c.f. Fig.~\ref{fig:TLSNoise}.
Our model for the origin of the fluctuations is then based on the presence of a large number of \emph{interacting} TLS at both low and high eigenfrequencies.
Due to the interactions between TLS, thermal switching of the state of low-frequency TLS will then lead to fluctuations in the energy splitting of high-frequency TLS, 
providing a qualitative description of the observed data.
This model is further underpinned by our direct observation of fluctuations in a high-frequency TLS' energy splitting, which occurs on time scales comparable with the qubit's $T_{1}$ fluctuation; 
see Fig.~\ref{fig:TLSEnergy}.
In the following, we will indicate TLS with eigenenergies much larger than the thermal energy as TS (tunnelling systems), while those at energies much lower than temperature will be named TF (thermal fluctuators).

Our model provides a qualitative description of the origin of fluctuations in the electrical susceptibility of mesoscopic circuits, 
an area which has recently started to attract attention from both experiment and theory~\cite{Sendelbach:PRL:2009, Neill:APL:2013, Schad:PRB:2014}.
We also note that interactions between TLS have recently been observed directly in two strongly coupled defects~\cite{Lisenfeld:NC:2015} and that such coupling has been invoked as a model of noise before, 
e.g. to explain the line-width broadening and spectral diffusion of ultrasonic excitations of TLS ensembles in glasses~\cite{Arnold:SSC:1972, Black:PRB:1977} 
as well as spectral blinking of dye molecules~\cite{Boiron:ChemPhys:1999} and quantum dots~\cite{Frantsuzov:PRL:2009}.
More recently, Refs.~\onlinecite{Faoro:PRB:2015, Burnett:NC:2014} make a connection between slow fluctuations in the resonance frequency of superconducting resonators causing phase noise, 
and ensembles of interacting TLS leading to fluctuations in the energy splitting of high-frequency TS, much along the same lines as we describe here.
While in that work the real part of the susceptibility was considered, leading to fluctuations in the level splitting of a resonator, here we are concerned with its imaginary part that is responsible for energy dissipation.

\section{Experimental evidence\label{sec:Exp}}
The fluctuations of the $T_{1}$-time reported here (Fig.~\ref{fig:Gamma1Time}) were measured in a superconducting qubit in the 3D-transmon design~\cite{Paik:PRL:2011}, 
with an average relaxation time $T_{1}$ of $\sim80~\mu$s. 
In our 3D-transmon circuit, the qubit energy, i.e. the level splitting of its two lowest levels, 
is fixed at $\omega_{q}/2\pi = 3.58$~GHz and not tuneable as in other designs~\cite{Makhlin:N:1998, Mooij:S:1999, Vion:S:2002, Riste:N:2013, Roch:PRL:2014}.
Each datapoint results from a series of individual measurements, each time resonantly exciting the qubit and detecting the qubit population after waiting for some time $t$. 
The resulting traces where fitted to an exponential decay curve $\propto \ee^{-\Gamma_{1} t}$.
The observed fluctuations of the qubit's relaxation rate $\Gamma_{1}$ do not show any apparent structure,
with the largest experimentally resolvable fluctuation rate given by the inverse of the time it takes to obtain a single value of $T_{1}$, here $\sim 1$~min.
Additional datasets are shown \ifthenelse{\boolean{app}}{in Appendix B}{in Ref.~\onlinecite{supplement}}.

In a second experiment, we use a superconducting phase qubit to directly monitor the properties of a single high-frequency TS that is strongly coupled to a superconducting phase qubit. Figure~\ref{fig:TLSEnergy}~(a) shows measured time-dependent fluctuations of the TS' energy level splitting which occur on time scales similar to those of the above discussed qubit fluctuations. Here, the TS' resonance frequency $E$ was repeatedly measured by varying the frequency of a long microwave pulse applied to the qubit circuit 
with a pulse amplitude that was large enough to allow for direct excitation of the TS excited state~\cite{Lisenfeld:PRL:2010}. 
During the microwave pulse, the qubit was kept far detuned from the TS. After the pulse, qubit and TS were brought into resonance in order to swap the TS excitation into the qubit, whose population was then read out. 
Details of this technique can be found in \ifthenelse{\boolean{app}}{Refs.~\onlinecite{Lisenfeld:PRL:2010, Lisenfeld:NC:2015} and Appendix A}
{Refs.~\onlinecite{Lisenfeld:PRL:2010. Lisenfeld:NC:2015, supplement}}.

\section{Theoretical Model\label{sec:Theo}}

\subsection{TLS as sources of fluctuating noise}
In the following, we describe our model explaining the observed fluctuations in the relaxation rate $\Gamma_{1} = 1/T_{1}$ of superconducting circuits. 
We first note that in a master equation description of dissipative quantum dynamics, the relaxation rate of a qubit is proportional 
to the unsymmetrized spectrum of its environment at the frequency of the qubit's level-splitting, 
$\Gamma_{1} \propto C(\omega_{q})$~\cite{Gardiner:2004}. Here we assume effectively zero temperature, $k_{B} T\ll \hbar \omega_{q}$, so that thermal excitations can be neglected.
It is then our goal to relate fluctuations in the energy of a  single TS to changes in the high-frequency noise acting on the superconducting circuit 
and to further characterise the fluctuations in terms of parameters of the experiments and the TLS distributions. 
We start by describing a single TLS as a quantum two-level system using the tunnelling Hamiltonian~\cite{Phillips:JLTP:1972, Anderson:PM:1972}
\begin{align}
	\H_{\text{TLS}} = -\frac12 \varepsilon \sigma_{z} + \frac12 \Delta \sigma_{x} \,,
	\label{eq:HTLS}
\end{align}
where $\varepsilon$ is the asymmetry energy between the two wells and $\Delta$ is the tunnel splitting. The Pauli-matrix $\sigma_{z}$ here describes the position of a particle on either side of a double-well potential, 
and $\sigma_{x}$ induces transitions between the wells. 
Diagonalizing yields $\H_{\text{TLS}} = \frac12 E \tilde{\sigma}_{z}$ with the level-splitting $E = \sqrt{\varepsilon^{2}+\Delta^{2}}$. Here and in the following we use the convention $\hbar =1$, 
so that all energies are expressed in units of angular frequencies.

The TS observed in high-frequency noise spectra are believed to be charged entities interacting with the superconducting circuits 
via their electric dipole moment $\propto \sigma_{z}$~\cite{Simmonds:PRL:2004}. 
Assuming weak qubit-TS coupling, their effect on the qubit will be that of strongly coloured noise, 
where the spectral density can be calculated from the Fourier transform of the two-time correlation function of their coupling operator $\sigma_{z}$~\cite{Shnirman:PRL:2005}. 
We obtain
\begin{align}
	C(\omega) &= \int dt\: \ee^{-\ii \omega t} \mean{\sigma_{z}(t) \sigma_{z}(0)} \nn \\
		&= \cos^{2}{\theta} \l[ 1 - \mean{\sigma_{z}}^{2} \r] \frac{ 2 \gamma_{1} }{ {\gamma_{1}}^{2} + \omega^{2} } \nn\\
		&+ \sin^{2}{\theta} \l[ \frac{ 1 + \mean{\sigma_{z}} }{2} \r] \frac{ 2 \gamma_{2} }{ {\gamma_{2}}^{2} + (\omega - E)^{2} } \nn\\
		&+ \sin^{2}{\theta} \l[ \frac{ 1 - \mean{\sigma_{z}} }{2} \r] \frac{ 2 \gamma_{2} }{ {\gamma_{2}}^{2} + (\omega + E)^{2} } \,,
	\label{eq:COmega}
\end{align}
with the TLS' equilibrium steady-state population $\mean{\sigma_{z}} = \tanh{(E / 2 k_{B} T)}$, the intrinsic TLS relaxation rate $\gamma_{1}$, and 
$\gamma_{2} = \frac12 \gamma_{1} + \gamma_{\varphi}$, where $\gamma_{\varphi}$ is the pure dephasing rate of the TLS. 
Here, $\tan{\theta} = \Delta/\varepsilon$ defines the TLS' mixing angle. 
Eq.~\eqref{eq:COmega} is composed of three parts, each of which is relevant for TLS in different parameter regimes. 
The first line describes low-frequency noise due to random switching of the TLS and is most pertinent for low-frequency TF with $E \ll k_{B} T$. 
The second term is a high-frequency contribution which is sharply peaked around the TLS energy and is most pronounced for TS with $E \gg k_{B}T$.
Since those TS are mostly resting in their ground state, they are able to absorb energy from the qubit.
It is this contribution that gives rise to the observed resonances in the noise spectrum~\cite{Astafiev:PRL:2004, Ithier:PRB:2005, Barends:PRL:2013} and in which we are mostly interested.
The final term contributes at negative frequencies and describes the ability of the TLS to excite the qubit by transferring an excitation to it. 
For both high-frequency TS in thermal equilibrium as well as low-frequency TF, this term will not contribute.

For simplicity, we assume the environmental noise at frequencies close to the qubit level splitting $\omega_{q}$ is dominated by a single, weakly coupled high-frequency TS at energy $E\sim\omega_{q}$.
We further assume this TS is interacting with a large number of other TLS which are located in its close spatial vicinity.
This is the situation illustrated in Fig.~\ref{fig:TLSNoise} and the one most relevant to experiment~\cite{Astafiev:PRL:2004, Ithier:PRB:2005, Barends:PRL:2013}. 
If the distribution of TS at high frequencies is dense~\cite{Shnirman:PRL:2005, Faoro:PRB:2015}, our results still hold but have to be additionally averaged over the high-frequency distribution. 
We model the interaction between all TLS in the sample by a Hamiltonian of the form
\begin{align}
	\H = \frac12 \tilde\sigma_{z}\: \sum_{j} g_{j} \tilde\sigma_{z,j} \,,
	\label{eq:TLSInt}
\end{align}
where $g_{j}$ is the coupling strength between the high-frequency TS and all other TLS, indicated by the index $j$. 
Coupling of the type Eq.~\eqref{eq:TLSInt} can be caused e.g. by electric dipole coupling or strain-mediated interaction, 
where the asymmetry-energy of either TLS depends on the relative position of the other TLS in their respective double-well potentials~\cite{Lisenfeld:NC:2015}. 
With such an interaction, the energy splitting of any TLS depends on the instantaneous state of all TLS in a certain range around it, 
determined by the microscopic origin of their interaction; c.f. inset to Fig.~\ref{fig:TLSNoise}.

	We are looking at fluctuations in the qubit relaxation rate due to slow fluctuations in the TS energies $E$. 
	In order to calculate expectation values and statistics, we write the level splitting of an individual TLS in the form
	$\op E = E_{0} - \sum_{j} g_{j} \tilde\sigma_{z,j}$,
	now depending on the state of all other TLS via the mutual interaction $ g_{j} $ from Eq.~\eqref{eq:TLSInt}.
	Here we focus on a high-frequency TS with $E_{0} \gg k_{B} T, \gamma_{2}$, 
	such that $\mean{\tilde\sigma_{z}} = -1$ and the resulting spectral density is strongly peaked around the TS eigenenergy $E$, c.f. Eq~\ifthenelse{\boolean{app}}{\eqref{eq:COmega}}{(2) in the main text}.
	We defined the undisturbed TS level splitting as $E_{0} = \sqrt{\varepsilon^{2} + \Delta^{2}}$
	with the parameters $\varepsilon$ and $\Delta$ from Eq.~\ifthenelse{\boolean{app}}{\eqref{eq:HTLS}}{(1) in the main text}.
	
	We can further write the qubit relaxation rate due to its coupling to a single high-frequency TS as 
	$\Gamma_{1} \propto \op\gamma_{q}$.
	Here, the relaxation rate induced by a single TS is given by the high-frequency components of its spectral density,
	\ifthenelse{\boolean{app}}{c.f. Eq.~\eqref{eq:COmega}}{c.f. Eq.~(2) in the main text}, as
	\begin{align}
		\op \gamma_{q} = \cos^{2}{\theta} \frac{2 \gamma_{2}}{ {\gamma_{2}}^{2} + (\omega_{q} - \op E)^{2}} \,.
	\end{align}
	Assuming the interaction between individual TLS to be weak, $g_{j} \ll \gamma_{2}$, we can expand this to first order as 
	\begin{align}
		\op \gamma_{q} = &\gamma_{q}^{(0)} + \gamma_{q}^{(1)} \sum_{j} g_{j} \tilde\sigma_{z,j} + O(g^{2}) \,,
		\label{eq:g10}
	\end{align}
	with the coefficients 
	\begin{align}
		\gamma_{q}^{(0)} &= \cos^{2}{\theta} \frac{ 2 \gamma_{2} }{ \gamma_{2}^{2} + (\omega_{q} - E_{0} )^{2} } \,, \label{eq:g11} \\
		\gamma_{q}^{(1)} &= \dvpart{\gamma_{q}}{E}\Bigr|_{E=E_{0}} 
			= \cos^{2}{\theta} \frac{4 \gamma_{2} (\omega_{q} - E_{0}) }{ \l({\gamma_{2}}^{2} + (\omega_{q} - E_{0})^{2} \r)^{2} }\,. \label{eq:g12}
	\end{align}
	Equations.~\eqref{eq:g10}-\eqref{eq:g12} will be the basis for our further calculations.

\subsection{Distribution of TLS parameters}	
	For tunneling TLS one usually assumes flat distributions for both the asymmetry energy $\varepsilon$ as well as the tunneling barrier height~\cite{Phillips:JLTP:1972, Anderson:PM:1972}.
	Since the tunneling energy $\Delta$ depends exponentially on the barrier, the resulting distribution in TLS parameters is $P(\varepsilon, \Delta) \sim 1/\Delta$. 
	The TLS relaxation rates are then also distributed log-uniformly, $P(\gamma_{1}) \sim 1/\gamma_{1}$, since the tunnelling strength depends mainly on the size of the tunnelling barrier.
	In Ref.~\onlinecite{Shnirman:PRL:2005} it was found that a linear or super-linear distribution in $\varepsilon$ 
	would naturally explain both low- and high-frequency parts of the noise spectrum acting on the qubit as stemming from the same ensemble of TLS.
	For the sake of generality, we will therefore assume the distribution of TLS parameters as 
	\begin{align}
		P(\varepsilon, \Delta) d\varepsilon d\Delta = A\: \frac{\varepsilon^{\alpha}}{\Delta} d\varepsilon d\Delta\,,
		\label{eq:PEpsilonDelta}
	\end{align}
	with $\alpha \geq 0$ and the constant $A$ needed for normalization. 
	For non-interacting TLS, the distribution is usually assumed to be flat, $\alpha = 0$~\cite{Phillips:JLTP:1972, Anderson:PM:1972},
	but might be different from zero in the more realistic case of interacting TLS~\cite{Shnirman:PRL:2005, Faoro:PRB:2015}.
	Without loss of generality we restrict the integration to the positive real axis. 
	The distribution of inter-TLS coupling strengths $g_{j}$ depends strongly on the physical model of their interaction. 
	It is important to note that the coupling strength $g$ in most models can be both positive or negative, meaning the coupling between the TLS can either raise or lower the energy of the respective partners. 
	For the dipole coupling model this reflects the fact that the relative orientation of the dipoles can be both parallel as well as antiparallel.		
	
\subsection{General considerations}	

In the calculations one has to carefully separate the different time scales of the problem. 
The measurement protocol fixes three distinct scales, which have to be compared to the fluctuation rates of individual low-frequency TF 
to determine the nature of their contribution to the fluctuations in the qubit's relaxation rate $\Gamma_{1}$. 
First, there is the time it takes to do a single measurement of the qubit population, $t_{\text{meas}}$, where many such measurements are averaged to obtain each point in a complete relaxation curve.
Fluctuating TF that are faster than $1 / t_{\text{meas}}$ will not contribute since they average out even for a single measurement.
Second, there is the time to measure a single point of a curve, $t_{\text{point}}$.
Fluctuations that are faster than $1 / t_{\text{point}}$, but slower than $1 / t_{\text{meas}}$ will act as an effective broadening of the high-frequency TS resonance, increasing its line-width $\gamma_{2}$.
The slowest time scale is given by the duration of the measurement of a complete $T_{1}$ curve, $t_{T_{1}}$. 
TF dynamics slower than $1/t_{\text{point}}$ but faster than $1/t_{T_{1}}$ will lead to jitter in the energy relaxation curve, contributing additional noise in the fit of $T_{1}$. 
Finally, slow TF that fluctuate at frequencies that are smaller than $1 / t_{T_{1}}$ will be the ones responsible for the low-frequency fluctuations visible in the $T_{1}$ data, see Fig.~\ref{fig:Gamma1Time}.
Note that the microscopic origin of these small switching rates is so far unclear~\cite{Faoro:PRB:2015}.
For the very slow fluctuations observed in experiments, on time scales $\sim$~min, to the best of our knowledge no microscopic model exists. 
A possible candidate might be collective behaviour of large ensembles of TLS that form clusters~\cite{Atalaya:PRB:2014, Kechedzhi:2011}, but clear experimental confirmation of this effect is missing so far.

In the following we will be interested in calculating the temperature and frequency dependence of the qubit relaxation rate due to its coupling to individual TS, averaged over TLS parameter distributions, 
as well as the spectrum of the fluctuations in $\Gamma_{1}$. 
Due to the considerations above, the temperature dependence will be strongly influenced by the thermally activated part of the TF distribution, contributing via the TS linewidth $\gamma_{2}$.
Following Refs.~\onlinecite{Shnirman:PRL:2005, Faoro:PRB:2015} we find the temperature dependence of the dephasing rate due to a bath of low-frequency TFs as $\gamma_{2} \propto T^{\alpha+1}$,
where $\alpha$ characterizes the TLS distribution.

\subsection{Average relaxation rate}	
	We now turn to calculating the average of the qubit's relaxation rate using the distributions introduced above. 
	We concentrate here on fluctuations originating from the low frequency contributions from TF with small level splitting, $E \lesssim k_{B} T$, 
	since those are the ones directly observable in experiment. 
	Noting that for TLS in thermal equilibrium $\mean{\tilde\sigma_{z}} = \cos{\theta} \mean{\sigma_{z}} = \cos{\theta} \tanh{(E / 2  k_{B}T)}$, 
	we can directly write down the mean value of the qubit relaxation rate due to the high-frequency TS to lowest order in the inter-TLS coupling strength $g$ as
	\begin{align}
		\mean{\op \gamma_{q}} &= \gamma_{q}^{(0)} + \gamma_{q}^{(1)} \sum_{j} g_{j} \cos{\theta_{j}} \tanh{\frac{E_{j}}{2 k_{B} T}} 
	\end{align}
	where the sum includes all other two-level defects that a single high-frequency TS interacts with.
	In the calculation of the average rate $\mean{\op \gamma_{q} }$, we immediately notice that
	$\int dg\: g P(g) = 0$,
	since we integrate an odd function over an even range. 
	Therefore we simply find 
	\begin{align}
		\mean{\op \gamma_{q}} = \gamma_{q}^{(0)} \,,
	\end{align}
	i.e., the average contribution to the relaxation rate from a single TS is given by it's spectrum centred around its undisturbed level-splitting $E_{0}$.	
For the temperature dependence of the ensemble-averaged qubit relaxation rate we then find
\begin{align}
	\mean{\Gamma_{1}} \propto \frac{ 2 \gamma_{2}}{ {\gamma_{2}}^{2} + \delta\omega^{2} } \propto 
	\begin{cases}
		T^{-(\alpha+1)} &\,,\quad \delta\omega \lesssim \gamma_{2} \\
		T^{\alpha+1} &\,,\quad \delta\omega \gg \gamma_{2} \\
	\end{cases}\,,
	\label{eq:GammaMean}
\end{align}
where $\delta\omega = \omega_{q} - E$ is the detuning between qubit and TS, and we distinguish between the case where qubit and TS are nearly resonant and when they are far detuned.

\subsection{Rate fluctuation spectrum}	
	The spectrum of fluctuations of the qubit relaxation rate is then related to the Fourier transform of the rate correlation function as
	\begin{align}
		\mean{ \op\gamma_{q}(t) \op\gamma_{q}(0) }_{\omega} &= \int dt\: \ee^{-\ii\omega t} \mean{ \op\gamma_{q}(t) \op\gamma_{q}(0) } \nn\\
			&= \l( \gamma_{q}^{(1)} \r)^{2} \sum_{j,l} g_{j} g_{l} \mean{\tilde\sigma_{z,j}(t) \tilde\sigma_{z,l}(0)}_{\omega} 
		\label{eq:ST1}
	\end{align}
	where, in evaluating the correlator, we restrict ourselves to the low frequency contribution of the TLS autocorrelation function Eq.\ifthenelse{\boolean{app}}{~\eqref{eq:COmega}}{~(2) in the main text},
	i.e. we focus on TFs with $E\ll k_{B} T$. 
	Additionally we assume that different TLS are uncorrelated, $\mean{\sigma_{z,j}\sigma_{z,l}} = 0$. 
	We are also only interested in the bare fluctuations of the rate, so we have already subtracted the mean value above. 
	For more details on the calculations, see Appendix C, where we additionally discuss the case when the TF switching is solely due to interactions with phonons.
	
	For the average over the coupling strength, one finds
	$\int dg\: g^{2} P(g) \propto \text{const}$,
	where the constant is mainly determined by the maximum possible coupling strength and thus by the minimal distance between TLS and the microscopic origin of their interaction. 
	Performing the average over the mixing angle $\theta$ also contributes a constant, with the exact value again depending on details of the microscopic TLS model.
	The average over TF energies can be written as 
	\begin{align}
		\int dE\: P(E) \l( 1- \tanh^{2}{\l( \frac{E}{2k_{B}T} \r)} \r) \approx \int_{0}^{T} dE\: E^{\alpha} = T^{\alpha+1} \,,
	\end{align}
	contributing to the temperature dependence of the final result.
	Still assuming small interaction strength between TLS, $g \ll \gamma_{2}$, we can now distinguish three regimes related to the initial detuning between our qubit and the high-frequency TS, 
	$\delta\omega = \omega_{q} - E_{0}$. 
	For qubit and high-frequency TLS near resonance, $\delta\omega \ll \gamma_{2}$, we find that $\gamma_{q}^{(1)} \propto \delta\omega / \gamma_{2}^{3}$, 
	while in the regime of intermediate detuning, $\delta\omega \sim \gamma_{2}$, 
	one finds $\gamma_{q}^{(1)} \propto 1 / \gamma_{2}^{2}$.
	In the far detuned regime, $\delta\omega \gg \gamma_{2}$, we finally have $\gamma_{q}^{(1)} \propto \gamma_{2} /\delta\omega^{3}$.
	Finally, adopting the standard assumption for tunnelling TLS, $P(\gamma_{1}) \sim 1/\gamma_{1}$, the frequency dependence of the fluctuation spectrum is determined by
	\begin{align}
		\int_{0}^{\gamma_{\text{Max}}} d\gamma_{1}\: P(\gamma_{1}) \frac{2\gamma_{1}}{\gamma_{1}^{2} + \omega^{2}} 
			\propto 
			\frac{1}{\omega} &\,,\quad \omega<\gamma_{\text{Max}} \,.
	\end{align}
	Here the maximum relevant switching rate $\gamma_{\text{Max}}$ is given by the time of a single $T_{1}$-measurement. All faster fluctuations will be averaged out in the observations, 
	leading to the behaviour $\sim 1/\omega$ for $\omega<\gamma_{\text{Max}}$.
	
Thus, we find the temperature and frequency dependence of the $T_{1}$ fluctuation-spectrum as
\begin{align}
	\mean{ \Gamma_{1}(t) \Gamma_{1}(0) }_{\omega} \propto \omega^{-1}
	\begin{cases}
		T^{-5(\alpha + 1)} &\,,\quad \delta\omega \ll \gamma_{2}\\
		T^{-3(\alpha + 1)} &\,,\quad \delta\omega\sim\gamma_{2}\\
		T^{2(\alpha + 1)} &\,,\quad \delta\omega\gg\gamma_{2} \\
	\end{cases} \,,
	\label{eq:GammaVar}
\end{align}
where $\mean{\ldots}_{\omega}$ denotes the Fourier transform of the two-time correlation function of the relaxation rate $\Gamma_{1}$.
These results hold for small inter-TLS coupling $g_{j}\ll\gamma_{2}$. The opposite case $g_{j}\gg\gamma_{2}$ corresponds to on-off switching and is excluded by the experimental data. 
In the intermediate regime, $g_{j}\sim\gamma_{2}$, the overall temperature dependence will be given as an average over our results.

\section{Discussion\label{sec:Disc}}

\subsection{Implications and tests of the model}
Our model can be directly tested by measuring the relaxation rate at different qubit level splittings and inferring the time and frequency dependence of the noise spectrum acting on the qubit. 
By using a frequency-tunable qubit, the fluctuations in the noise spectral density might be directly resolvable in time and frequency, 
depending on the time-scale of a single measurement of the relaxation time $T_{1}$. 
Even for non-tuneable qubits it is possible to probe the noise spectral density in close vicinity of the qubit frequency 
by measuring the decay of Rabi oscillations of the qubit\ifthenelse{\boolean{app}}{, c.f. Appendix D and Ref.~\onlinecite{Hauss:NJP:2008}}{~\cite{Hauss:NJP:2008, supplement}}.
Another possibility is to apply external driving to saturate the TF responsible for the fluctuations in TS energy. 
If an electric field is applied resonantly with the relevant low-frequency TF, it will lead to oscillations 
with the Rabi frequency depending on the detuning between drive tone and TF energies, the TF dipole moments and the electric field strength at their position.
Assuming the resulting Rabi frequency is fast compared to the duration of a single $T_{1}$ measurement, 
the effect would be to raise the average $\mean{\Gamma_{1}}$ while at the same time reducing the amplitude of its fluctuations. 
This is because the resonant driving of initially very slow TF will alter their contribution towards a simple line-width broadening of the high-frequency TS.
In experiments with 3D-transmon qubits this could be achieved by careful engineering of the cavity modes, 
such that there exists a suitable low-frequency mode exhibiting strong electric field components spatially close to the qubit. 
In other qubit architectures this might be possible within existing experimental setups~\cite{Lisenfeld:PRL:2010}. 
In our transmon qubit sample, this experiment proved unfeasible due to design restrictions in the employed cavity.
Additional verification could be achieved by a systematic characterisation of the fluctuations of $T_{1}$ at a variety of experimental temperatures $T$. 
An additional challenge arrises from the fact that the exact temperature dependence is connected sensitively to the qubit-TS detuning $\delta\omega$, c.f. Eq.~\eqref{eq:GammaVar}, 
which also has to be determined in this case.

\subsection{Alternative models}
Possible alternative models for the fluctuating noise spectrum include fluctuations of the quasiparticle density in the superconductor. 
Quasiparticle tunnelling across the circuit's Josephson junctions can induce relaxation and dephasing~\cite{Catelani:PRB:2011}, 
and explains well the temperature dependence of qubit relaxation rates for elevated sample temperatures. 
In contrast to our model, which depends on a structured noise spectrum as background, the quasiparticle induced noise is flat at high-frequencies. 
Following Ref.~\onlinecite{Catelani:PRB:2011} we calculate the fluctuations in quasiparticle density required to effect the observed variance in the relaxation time of transmon qubits.
For the parameters of our sample, we find the fluctuation in the quasiparticle volume density required to change the relaxation rate by $1$~kHz as 
$\delta n_{qp} \approx 0.5/\mu\text{m}^{3}$\ifthenelse{\boolean{app}}{, see Appendix E for details}{~\cite{supplement}}.
From the geometry of our sample, it then follows that this change would require the number of quasiparticles present on either one of the qubit islands to fluctuate by $\delta N_{qp} \approx 1.5 \times 10^{4}$.
We are not aware of any mechanism leading to symmetric fluctuations in the quasiparticle number of this magnitude.

Another possible model is that in the 3D-transmon sample used to obtain Fig.~\ref{fig:Gamma1Time}, the qubit level splitting might fluctuate in time, 
e.g. due to changes in the critical current of the circuits Josephson junction~\cite{Nugroho:APL:2013, Murch:APL:2012}. 
Together with the observed strong structure in the noise spectrum~\cite{Astafiev:PRL:2004, Ithier:PRB:2005, Barends:PRL:2013} this could also explain the fluctuations in the qubit relaxation.
Here we again have to be mindful of the time scales involved. Fast fluctuations of the qubit energy, i.e. faster than the Rabi frequency used to excite it (in our experiments $\Omega/2\pi \sim 4$~MHz), 
will not lead to the observed slow fluctuations in the relaxation rate, but rather average out over the measurement time $t_{T_{1}}$. 
Their effect would be such that the observed qubit relaxation rate would no longer depend on the noise spectrum at a single frequency, but rather an average over the spectrum at a range of frequencies.
Intermediate energy fluctuations, faster than $1/t_{T_{1}}$ but slower than the Rabi frequency, will lead to a different qubit energy at each point of a single measurement and thus manifest
as an additional source of noise in the fit parameters.
Very slow fluctuations in the qubit energy, on the same time scales as the variations in $T_{1}$ will, in addition to potentially impacting the qubit relaxation rate, also influence the resonance condition for the 
Rabi pulse used to excite the circuit for each measurement. 
From our experimental data we obtain the excitation amplitude as a fit parameter for each measurement, c.f. Appendix B. 
While there are fluctuations visible in these parameters, they are generally of small amplitude and not highly correlated with the observed $T_{1}$ variations. 
We therefore conclude that while this mechanism might be present, its effect is likely to be smaller than the one associated with fluctuating TLS. 

\section{Conclusion}
In this paper we present a simple model of interacting TLS which offers a qualitative understanding of the observed fluctuations in the relaxation times $T_{1}$ of superconducting quantum circuits.
The model is grounded in our experimental observations, grants a clear route towards further confirmation, and provides a way to verify and refine the existing microscopic TLS models.
Moreover, our model clearly indicates that parasitic TLS are a limiting factor for the stability of today's best performing superconducting circuits.
A better understanding of this decoherence source is thus vital for further improving the fidelity of superconducting quantum circuits.

\begin{acknowledgments}
	For the samples used in this work, we thank G.A.~Keefe and M.B.~Rothwell for fabricating the 3D-transmon qubit and J.~M.~Martinis for providing the phase qubit sample. 
	We thank A.~Blais, J.~Clarke, J.H.~Cole, M.~Marthaler, T.M.~Stace and M.~Steffen 
	for valuable comments and discussions. 
	We acknowledge financial support by the Army Research Office through the Intelligence Advance Research Program (IARPA) under Contract No. 
	W911NF-10-1-0324, German Research Foundation (DFG Grants LI 2446/1-1 and SCHO 287/7-1 / SH 81/2-1), 
	German-Israeli Foundation (GIF Grant 1183-229.14/2011).  Theory analysis was supported by 
	Russian Science Foundation (RNF Grant No. 14-42-00044). 
	CM acknowledges the support of the RMIT Foundation through an International Research Exchange Fellowship.
\end{acknowledgments}
		
\bibliographystyle{apsrev4-1}
\bibliography{T1Fluctuations}	

\newpage
\setcounter{figure}{0}
\setcounter{equation}{0}
\makeatletter 
\renewcommand{\thefigure}{A\arabic{figure}}
\renewcommand{\theequation}{A.\arabic{equation}}
\ifthenelse{\boolean{app}}{}{\renewcommand{\thesection}{S.\arabic{section}}}

\ifthenelse{\boolean{app}}{\section*{Appendix A: Sample parameters and experimental remarks}}{\section{Sample parameters and experimental remarks}}
	
\subsection{Observation of $T_{1}$-fluctuations of a 3D-transmon}
	The three-dimensional cavity resonator used in this work was machined from bulk aluminium 6061. 
	The cavity has a nominal size of $18.6~\text{mm} \times 15.5~\text{mm} \times 4.2~\text{mm}$ engineered to give a resonant frequency of approximately $12$~GHz. 
	Two bulk head SMA connectors are used as input and output ports. 
	The loaded quality factor of the waveguide cavity is $3\times10^{4}$, with the output connector stronger coupled than the input connector in order to guarantee a high signal-to-noise ratio. 
	The sample is shielded with a cryoperm can that is thermally anchored to the mixing chamber of the dilution refrigerator.
	
	The qubit manipulation and readout pulses are delivered to the cavity via a single coax line filtered by a 10 dB attenuator at each temperature stage of the refrigerator. 
	Input and output ports are directly connected to SMA filters made of eccosorb in order to block infrared radiation and thermalize the central conductor of the coupling connectors.
	
	The output readout signal is amplified by a chain of cryogenic and room temperature amplifiers for a total gain of 60 dB. 
	A low pass filter and two cryogenic circulators are used between the sample and the cryogenic amplifier. 
	The qubit state is readout via the dispersive shift of the waveguide cavity~\cite{Wallraff:PRL:2005}.
	
	The qubit is fabricated on sapphire substrate via aluminium double angle evaporation. 
	Two rectangular pads of $350~\mu\text{m} \times 700~\mu\text{m}$, separated by $50~\mu$m, 
	are connected by a $1~\mu$m wide aluminium strip with a Josephson junction of size $0.1~\mu\text{m} \times 0.1~\mu\text{m}$. 
	The chip has a total size of $3.0~\text{mm} \times 6.7~\text{mm}$  and is kept in place in the waveguide cavity by small pieces of indium. 
	The qubit is placed at the maximum of the electric field of the first cavity mode.
	
	The qubit used in this work has an energy gap $\omega_{q}/2\pi$ of 3.5825~GHz, and anharmonicity ($\sim E_C/2\pi$) of $171$~MHz.
	The qubit is designed to work in the transmon regime, with $E_J/E_C \sim 61$~\cite{Koch:PRA:2007}.
	
	For the $T_{1}$ measurements reported here, the qubit was resonantly excited with a microwave pulse amplitude leading to the Rabi frequency of $\Omega/2\pi = 3.5714$~MHz, 
	corresponding to a $\pi$-pulse duration of $140$~ns. 
	
\subsection{Observation of TLS frequency fluctuations}	
	Our theoretical model to explain time-dependent fluctuations in the energy relaxation time of superconducting qubits is based on their near-resonant coupling to high-frequency TLS, 
	with the additional assumption that those TLS themselves experience resonance frequency variations due to their interaction with thermally fluctuating defects at low frequencies. 
	In this work, we include first experimental evidence that individual high-frequency TLS may indeed show resonance frequency fluctuations in time as shown in 
	\ifthenelse{\boolean{app}}{Fig.~\ref{fig:TLSEnergy}.}{Fig.~3 in the main paper.}
	
	In order to access TLS individually, we exploit their strong coupling to the state of a superconducting phase qubit when they are residing in the amorphous tunnel barrier of the qubit's Josephson junction. 
	We were using a phase qubit sample that has been developed in the group of Prof. J. Martinis at University of California, Santa Barbara, USA, with sample parameters as described in Ref.~\onlinecite{Steffen:PRL:2006}.
	
	We recorded the Lorentzian resonance curve of the TLS by varying the frequency of a long microwave pulse applied to the circuit while the qubit was kept far detuned from the TLS. 
	As described in Ref.~\onlinecite{Lisenfeld:PRL:2010}, this allows one to resonantly drive TLS while they remain effectively decoupled from the qubit dynamics. 
	To read out the TLS quantum state, the qubit is first prepared in its ground state and then tuned into the TLS resonance. 
	This realises an iSWAP operation that maps the TLS state onto the qubit, where it can be measured.
	
	Some of the TLS that were investigated with this method showed time-dependent fluctuations of their resonance frequency that were large enough to be resolved spectroscopically. 
	Often, we observe telegraph-signal like switching of TLS resonance frequencies between two similar values, indicating coupling to one dominating thermally activated TLS at low frequency. 
	
	To characterise the internal TLS parameters, tunnelling energy $\mathit{\Delta}$ and asymmetry energy $\varepsilon$ were measured 
	by recording the strain dependence~\cite{Grabovskij:S:2012} of its resonance frequency 
	and performing a hyperbolic fit to the equation $E = \sqrt{\mathit{\Delta}^2+\varepsilon^2}$. 
	\ifthenelse{\boolean{app}}{Figure~\ref{fig:TLSEnergy}}{Figure~3 in the main text}
	was obtained on a TLS that had $\mathit{\Delta} / 2\pi = 7.056$ GHz and whose asymmetry energy was tuned to $\varepsilon / 2\pi = 918$ MHz. 
	At this asymmetry, this TLS had an energy relaxation time of $T_1 \approx 590$ ns and a dephasing time of about $T_2 \approx 500$ ns. The sample temperature was kept at 33 mK.	

\ifthenelse{\boolean{app}}{\section*{Appendix B: Additional Experimental Data}}{\section{Additional Experimental Data}}

	\begin{figure}[htbp]
		\begin{center}
			\includegraphics[width=.9\columnwidth]{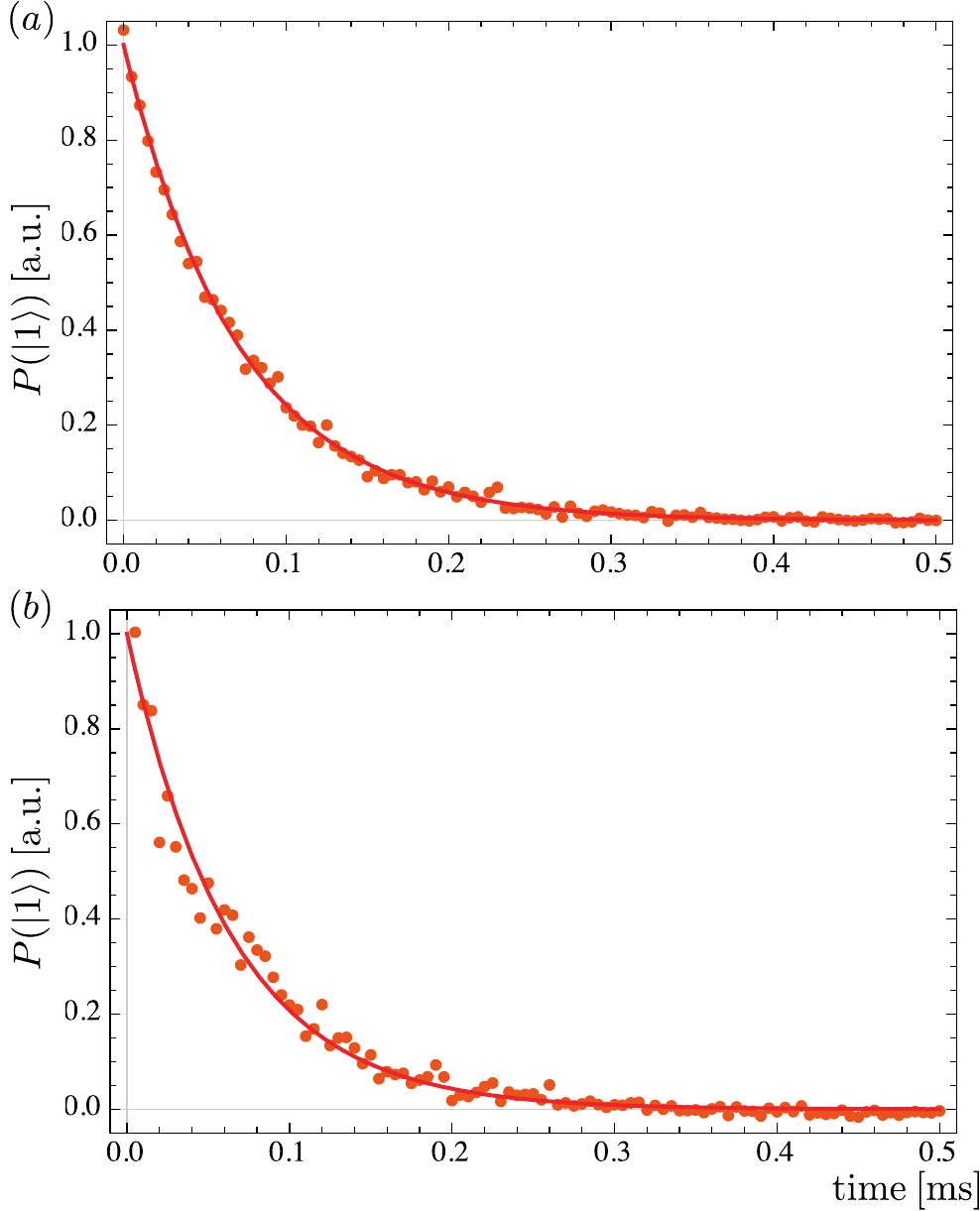}
			\caption{(Color online) Examples of decay curves taken at 30~mK. Points are experimental data of the qubit excitation probability $P(\ket 1)$ as a function of time after an initial $\pi$-pulse applied at $t=0$,
				normalised to lie between 0 and 1. The red curve is the result of a fit of the data to the function $P(\ket 1) = A \ee^{-\Gamma_{1} t } + B$, with free parameters $A$, $B$ and $\Gamma_{1}$.
				We show one curve with minimum standard error in the decay amplitude $A-B$ (a) and another curve with maximum error (b). 
				The lower curve might be better described not by purely exponential decay, if for example during measurement of the data the noise spectrum shows a sudden jump.
			}
			\label{fig:RelaxCurves}
		\end{center}
	\end{figure}
	Figure~\ref{fig:RelaxCurves} shows two examples of measured relaxation curves of the 3D-transmon qubit and the fits to the data.
	We fit the measurements to decay curves of the form $A \ee^{-\Gamma_{1} t} + B$ with the free parameters $A$, $B$ and $\Gamma_{1}$. 
	We show one trace where the fit converged with a very small standard error (a) and another where the convergence was worse (b). 
	The second trace might be better fit by assuming a double exponential decay where at some time the decay rate changed spontaneously 
	due to a change in the environmental noise spectrum (not shown).
	
	\begin{figure}[htbp]
		\begin{center}
			\includegraphics[width=.9\columnwidth]{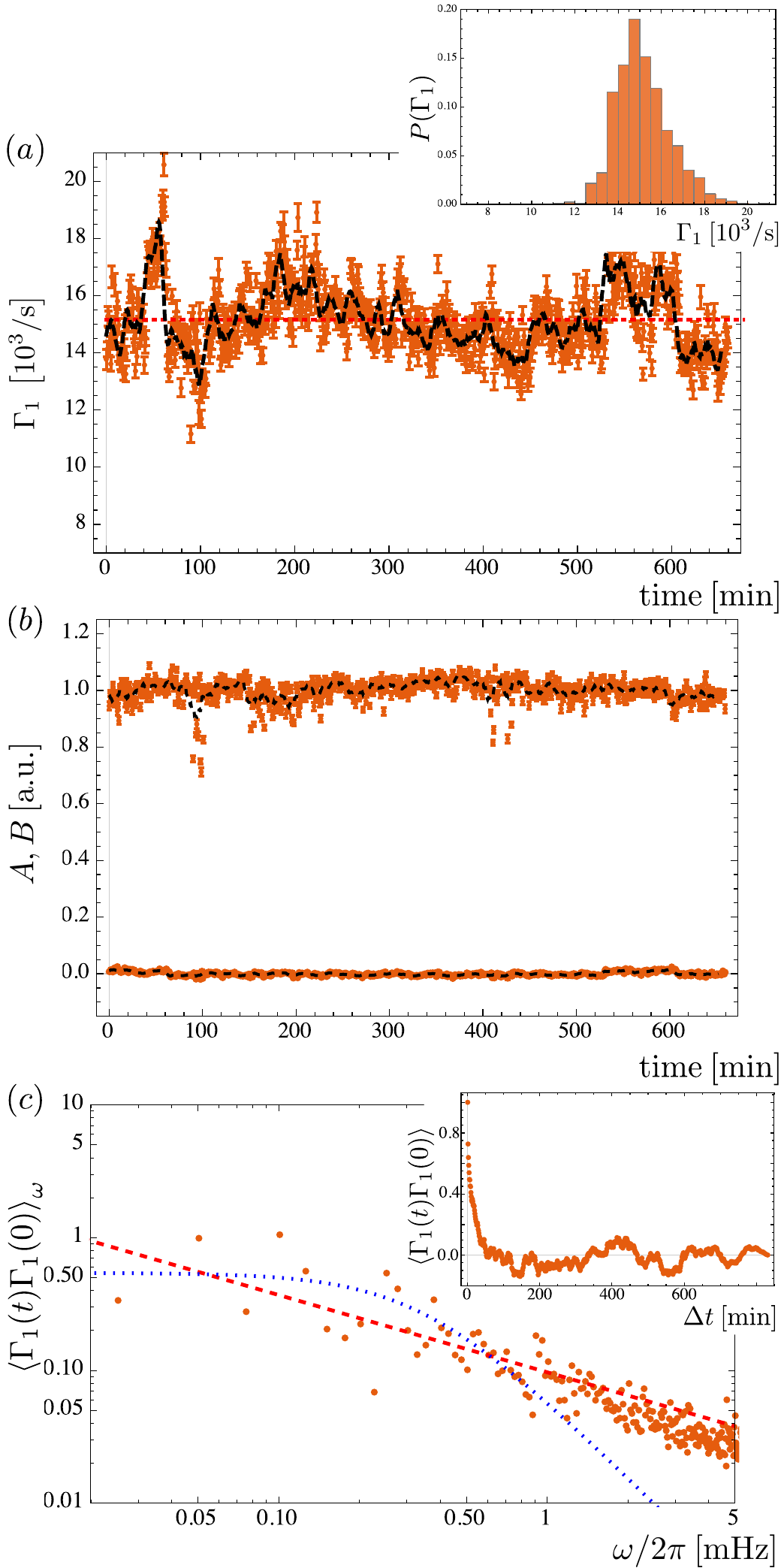}
			\caption{(Color online) Experimental data on $T_{1}$ fluctuations in the 3D-transmon sample at a temperature of $30$~mK. 
				\textbf{(a)} shows the relaxation rates $\Gamma_{1}$ from fits of the experiments to an exponential decay curve, $P(\ket 1) = A \ee^{-\Gamma_{1} t} + B$, 
				with error bars corresponding to the 95\% confidence interval of the fits. 
				The black dashed lines are a moving average over 10 points and the red dotted lines are the mean values over the full dataset. 
				The inset shows a histogram of the probabilities of values for the relaxation rate $\Gamma_{1}$.
				\textbf{(b)} shows the time evolution of excitation amplitudes $A$ and background $B$ from the same fits, including error bars and moving averages in black. 
				\textbf{(c)} depicts the absolute value of the Fourier transform of the two-time correlation function of the relaxation rates $\mean{\Gamma_{1}(t) \Gamma_{1}(0)}$ in a log-log-plot, 
				with the inset showing the correlation function itself. 
				The red (blue) dashed curve is the result of a fit of the data to a $A/\omega^{\alpha}$-spectrum [Lorentzian spectrum $A \gamma / (\gamma^{2} + \omega^{2})$]
				with fit parameters $A = 0.097$ and $\alpha = 0.58$ [$A = 0.18$ and $\gamma = 0.34$~mHz], for details see text.
			}
			\label{fig:Data30mK}
		\end{center}
	\end{figure}
\begin{figure}[htbp]
		\begin{center}
			\includegraphics[width=.9\columnwidth]{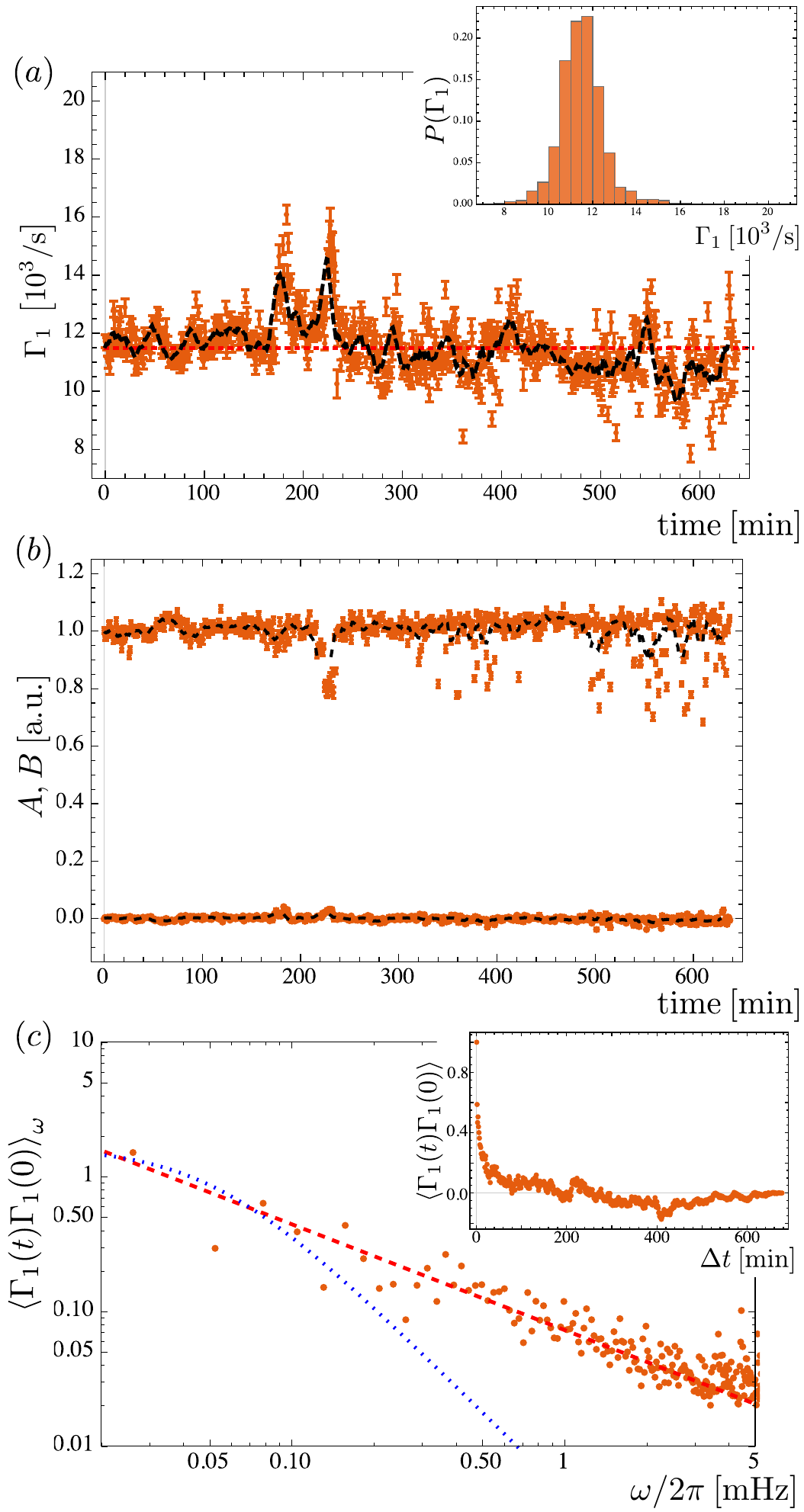}
			\caption{(Color online) Same as Fig.~\ref{fig:Data30mK}, data taken from experiments performed at a temperature of 50mK.
				Fit parameters in (c) are $A=0.079$ and $\alpha = 0.79$ [$A=0.087$ and $\gamma=0.052$~mHz] for red (blue) dashed line.
			}
			\label{fig:Data50mK}
		\end{center}
	\end{figure}
\begin{figure}[htbp]
		\begin{center}
			\includegraphics[width=.9\columnwidth]{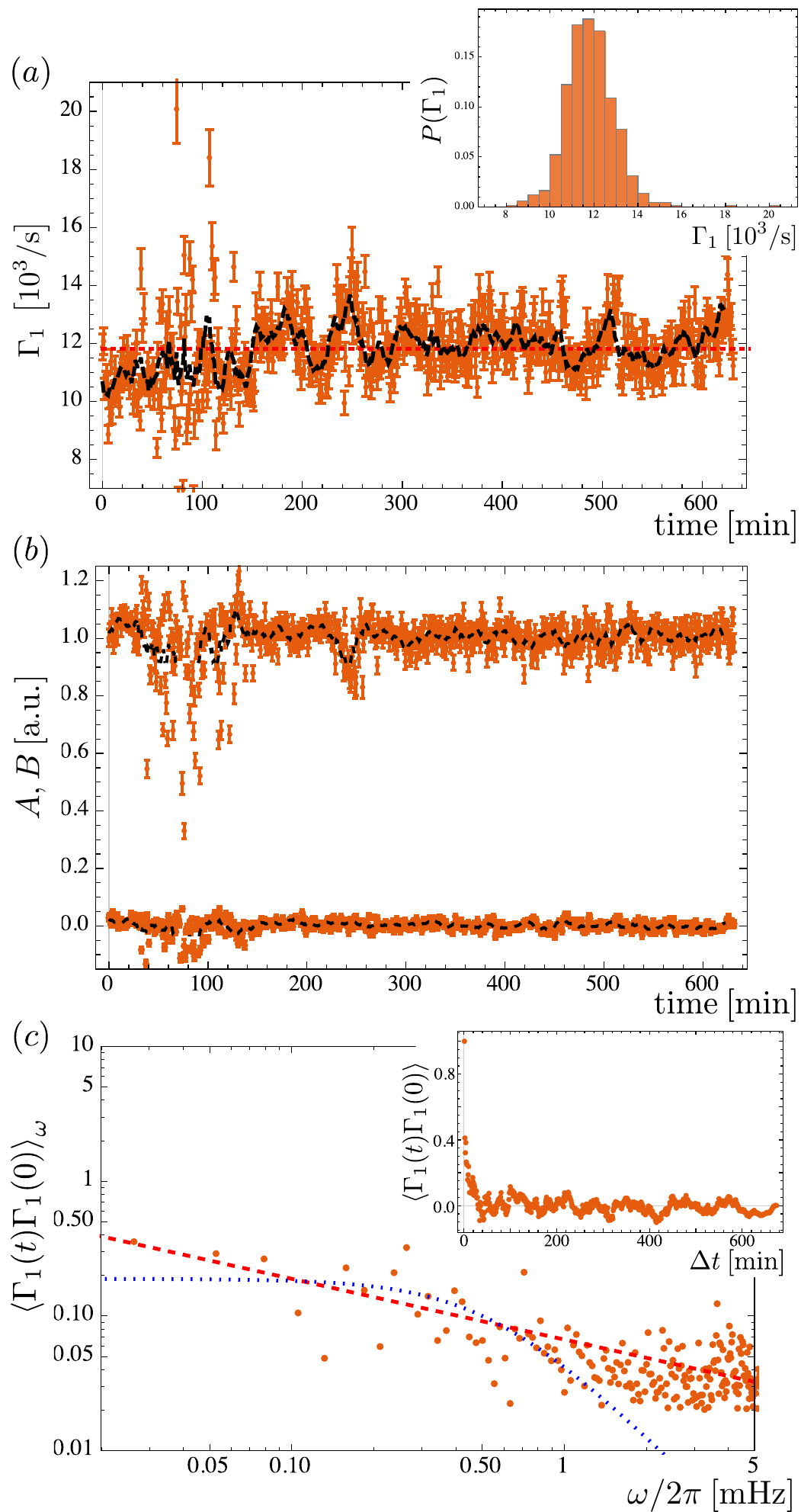}
			\caption{(Color online) Same as Fig.~\ref{fig:Data30mK}, for a sample temperature of 100mK.
				Fit parameters in (c) are $A=0.067$ and $\alpha = 0.45$ [$A=0.36$ and $\gamma=4.91$~mHz] for red (blue) dashed line.
			}
			\label{fig:Data100mK}
		\end{center}
	\end{figure}

	Figs.~\ref{fig:Data30mK}-\ref{fig:Data100mK} show the full datasets of the fluctuations in the relaxation rate $\Gamma_{1}$ measured in our 3D-transmon at three different experimental temperatures. 
	We also show the histograms for the probability of occurrence of a particular value of $\Gamma_{1}$ for all three temperatures as well as the fluctuations in the fit amplitude $A$ and background $B$. 
	The later two show some fluctuations, but are relatively flat on the scale of the changes observed in $\Gamma_{1}$. 
	Amplitude fluctuations might be explained if the qubit's level splitting varies in time, which, 
	together with a strongly coloured high-frequency noise spectrum provides an alternative model for the fluctuations in the qubit's relaxation rate (c.f. main text).
	From the data in Figs.~\ref{fig:Data30mK}-\ref{fig:Data100mK}, we conclude that this mechanism might be present but is weak and not the main contribution. 
	Additionally, we show the two-time correlation function of the relaxation rate as well as its Fourier transform. We fit the $T_{1}$-fluctuation spectrum 
	to two different functions and show the results in the plots.
	The red dashed lines are from the best fit to the function $A/\omega^{\alpha}$, corresponding to a $1/f$-type frequency distribution 
	as it is expected from a dense distribution of low-frequency TLS~\cite{Dutta:RMP:1981,Shnirman:PRL:2005}
	The blue dashed lines are results from a fit to a zero-frequency Lorentzian $\sim A \gamma / ( \gamma^{2} + \omega^{2})$, 
	as it would result from a single dominant low-frequency TLS, c.f. Eq.\ifthenelse{\boolean{app}}{~\eqref{eq:COmega}}{~(2) in the main text}.
	For our data presented here, the temperature dependence of the fluctuation amplitude is inconclusive and does not give any indication if our model is accurate.
	On the other hand, the frequency dependence of the correlations seems to follow roughly a $1/\omega$ dependence, which can be explained in the terms of our model.
	
\ifthenelse{\boolean{app}}{\section*{Appendix C: Model Calculations}}{\section{Calculations of the model}}

	Here we give additional details on the calculations of the mean value and spectrum of the $T_{1}$-fluctuations in a superconducting circuit due to interactions within a bath of spurious background TLS.
		
\subsection*{TLS parameter distribution}	
	Rewriting Eq.~\eqref{eq:PEpsilonDelta} in terms of the TLS level-splitting $E$ and the mixing angle $\theta = \arctan{\Delta/\varepsilon}$, we find
	\begin{align}
		P(E, \theta) d E d\theta= A\: E^{\alpha} \frac{\cos^{\alpha}{\theta} }{\sin{\theta} } d E d\theta\,.
	\end{align}
	When describing the full distribution of TLS for all energies, we integrate the tunnel splitting $\Delta$ between $\Delta_{\text{Min}} \gtrsim 0 $ and $\Delta_{\text{Max}}$ 
	and the asymmetry energy $\varepsilon$ between $\varepsilon_{\text{Min}} = 0$ and $\varepsilon_{\text{Max}}$. 
	We find for the integration bounds in the new variables: $\theta_{\text{Min}} = \arctan{\Delta_{\text{Min}} / \varepsilon_{\text{Max}} } \gtrsim 0$, 
	$\theta_{\text{Max}} = \arctan{\Delta_{\text{Max}} / \varepsilon_{\text{Min}}} = \pi / 2$ and
	$E_{\text{Min}} = \sqrt{\Delta_{\text{Min}}^{2} + \varepsilon_{\text{Min}}^{2} } = \Delta_{\text{Min}}$, $E_{\text{Max}} = \sqrt{ \Delta_{\text{Max}}^{2} + \varepsilon_{\text{Max}}^{2}}$.
	Here, $\Delta_{\text{Min}}$ is defined by the minimum tunneling barrier below which the description as a two-level system breaks down 
	and $E_{\text{Max}}$ provides an upper bound on the TLS level-splitting. 
	
	As an example for the distribution of the inter-TLS coupling strength $g$, 
	we write the probability distribution in the case where the interaction is mediated by dipolar interaction with $\abs{g} \sim 1/r^{3}$. 
	One finds 
	\begin{align}
		P(g) dg = P(r) \dvpart{r}{g} dg = \rho_{0} \abs{g}^{-\frac{4}{3}} dg \,,
	\end{align}
	where we assumed a constant TLS density in space $\rho_{0}$.

\subsection*{Calculating the average}	
	We here give some of the intermediate steps of the calculations of the average qubit relaxation rate and rate fluctuations spectrum.
	
	With the thermal occupation of a TLS in equilibrium, $\mean{\tilde\sigma_{z}} = \cos{\theta} \mean{\sigma_{z}} = \cos{\theta} \tanh{(E / 2 k_{B }T)}$, 
	we can directly write the mean value of the qubit relaxation rate to lowest order in the inter-TLS coupling strength $g$ as
	\begin{align}
		\mean{\op \gamma_{q}} &= \gamma_{q}^{(0)} + \gamma_{q}^{(1)} \sum_{j} g_{j} \cos{\theta_{j}} \tanh{\frac{E_{j}}{2 k_{B}T}} \nn\\
			&= \gamma_{q}^{(0)} + \gamma_{q}^{(1)} \int dg\: d\theta\: dE\: P(g,\theta,E) g \cos{\theta} \tanh{\frac{E}{2 k_{B}T}} \,,
	\end{align}
	where the sum includes all other two-level defects that a single high-frequency TS is interacting with. Due to the symmetric distribution in inter-TLS coupling strength $g$, 
	the above integral will evaluate to zero. 
	
	For the spectrum of fluctuations of the relaxation rate we then calculate the Fourier transform of the rate correlation function as
	\begin{widetext}
	\begin{align}
		\mean{ \op\gamma_{q}(t) \op\gamma_{q}(0) }_{\omega} &= \int dt\: \ee^{-\ii\omega t} \mean{ \op\gamma_{q}(t) \op\gamma_{q}(0) } \nn\\
			&= \l( \gamma_{q}^{(1)} \r)^{2} \sum_{j,l} g_{j} g_{l} \mean{\tilde\sigma_{z,j}(t) \tilde\sigma_{z,l}(0)}_{\omega} \nn\\
			&= \l( \gamma_{q}^{(1)} \r)^{2} \sum_{j} g_{j}^{2} \cos^{2}\theta_{j} 
				\l[ 1-\tanh^{2}\l( \frac{E_{j}}{2k_{B}T} \r) \r] \frac{ 2 \gamma_{1,j} }{ {\gamma_{1,j}}^{2} + \omega^{2} } \nn\\
			&= \l( \gamma_{q}^{(1)} \r)^{2} \int dg\: d\theta\: dE\: d\gamma_{1}\: P(g,\theta,E,\gamma_{1}) g^{2} \cos^{2}\theta
				\l[ 1-\tanh^{2}\l( \frac{E}{2k_{B}T} \r) \r] \frac{ 2 \gamma_{1} }{ {\gamma_{1}}^{2} + \omega^{2} } \,,
		\label{eq:ST1}
	\end{align}
	\end{widetext}
	where, in evaluating the correlator, we restrict ourselves to only the low frequency contribution of the TLS autocorrelation function Eq.\ifthenelse{\boolean{app}}{~\eqref{eq:COmega}}{~(2) in the main text},
	i.e. we focus on TFs with $E\ll k_{B}T$. 
	Additionally we are assuming that different TLS are uncorrelated, $\mean{\sigma_{z,j}\sigma_{z,l}} = 0$ and, since we are only interested in the bare fluctuations of the rate, 
	we have already subtracted the mean rate.
		
	Finally, adopting the standard assumption for tunnelling TLS, $P(\gamma_{1}) \sim 1/\gamma_{1}$, the frequency dependence of the fluctuation spectrum is determined by
	\begin{align}
		&\int_{0}^{\gamma_{\text{Max}}} d\gamma_{1}\: P(\gamma_{1}) \frac{2\gamma_{1}}{\gamma_{1}^{2} + \omega^{2}} \nn\\
			&\quad= \frac{2 \arctan{\frac{\gamma_{\text{Max}}}{\omega}}}{\omega}
			\propto 
			\begin{cases}
				\frac{1}{\omega} &\,,\quad \omega<\gamma_{\text{Max}} \\
				\frac{\gamma_{\text{Max}}}{\omega^{2}} &\,,\quad \omega>\gamma_{\text{Max}} \\
			\end{cases}\,.
	\end{align}
	Here the maximum relevant switching rate $\gamma_{\text{Max}}$ is given by the time of a single $T_{1}$-measurement. All faster fluctuations will be averaged out in the observations, 
	leading to the behaviour $\sim 1/\omega$ for $\omega<\gamma_{\text{Max}}$.
	In the opposite case $\omega \gg \gamma_{\text{Max}}$, i.e. when we observe the fluctuations on time scales that are short compared to $1/\gamma_{\text{Max}}$, 
	the spectrum will show a $1/\omega^{2}$ dependence.\\
	
\subsection*{Phonon induced TS switching}
	Alternatively to the generic thermally activated switching mechanism discussed previously,
	one can assume a microscopic model for the TLS relaxation rate $\gamma_{1}$. For example for coupling to phonons, and omitting irrelevant prefactors, one arrives at~\cite{Phillips:JLTP:1972}
	\begin{align}
		\gamma_{1} \propto \Delta^{2} E \coth{\l(\frac{E}{2k_{B}T}\r)} \propto  2 T E^{2} \sin^{2}{\theta} \,,
		\label{eq:GammaPhonons}
	\end{align}
	where in the second step we already assumed that the relevant energies of the switching TF are smaller than temperature, $E \ll T$.
	Since the relaxation rate in this expression depends mainly on the TF mixing angle $\theta$, the restriction on small switching rates will be realised by confining $\theta$ to small values around zero, 
	effectively limiting the value of the coupling strength between the relevant low-frequency TF and their phonon bath.
	Physically, Eq.~\eqref{eq:GammaPhonons} implies that phonons do not induce switching in TLS with small tunnelling matrix element $\Delta$.
	Performing the energy integration in the average we then get
	\begin{align}
		\int \:dE P(E) &\l( 1-\tanh^{2}{\l( \frac{E}{2T} \r)} \r) \frac{2\gamma_{1}}{\gamma_{1}^{2}+\omega^{2}} \nn\\
		&\approx \int_{0}^{T} \:dE E^{\alpha} \frac{4 T E^{2} \sin^{2}{\theta}}{4 T^{2} E^{4} \sin^{4}{\theta} + \omega^{2}} \nn\\
		&\sim \frac{ 4 T^{\alpha+4} \sin^{2}{\theta} }{(3+\alpha)\omega^{2}} \,,
		\label{eq:EPhonons}
	\end{align}
	where we expanded the integral to leading power in temperature $T$. 
	Combined with the previous results for the prefactor $\gamma_{q}^{(1)}$, this leads to the overall temperature and frequency dependence of the relaxation rate correlator
	\begin{align}
		\mean{\op\gamma_{q,i}(0) \op\gamma_{q,i}(t)}_{\omega} \propto \omega^{-2}
		\begin{cases}
			T^{-5\alpha - 2 } &\,,\quad \delta\omega \ll \gamma_{2,i}\\
			T^{-3\alpha} &\,,\quad \delta\omega\sim\gamma_{2,i} \\
			T^{2\alpha + 5} &\,,\quad \delta\omega\gg\gamma_{2,i} \\
		\end{cases}\,.
	\end{align}

	For the remaining integration over the mixing angle, one finds
	\begin{align}
		\int_{0}^{\theta_{\text{Max}}} \:d\theta P(\theta) \sin^{2}{\theta} &= \int_{0}^{\theta_{\text{Max}}} \:d\theta \sin{\theta}\cos^{\alpha}{\theta} \nn\\
			&= \frac{1-\cos^{1+\alpha}{\theta_{\text{Max}}}}{1+\alpha} \,,
	\end{align}
	where $\theta_{\text{Max}}$ is determined from Eq.~\eqref{eq:GammaPhonons} and the value of the maximum observable switching rate $\gamma_{\text{Max}}$. 
	In this case the temperature and frequency dependence to leading order in temperature is entirely contained in Eq.~\eqref{eq:EPhonons}.

\subsection*{Effective inter-TLS interaction range}
	Here we give a rough estimate of the maximum inter-TLS distance which still allows noticeable interactions between them.  
	We assume the TLS to be realised as microscopic electric dipoles of uniform dipole size $d_{i} = 1 e\times10^{-10}$~m, 
	where $e$ is the charge of a single electron~\cite{Simmonds:PRL:2004, Cole:APL:2010}.
	Then, assuming parallel orientation of the two TLS and using the relation between dipole magnitude and coupling strength
	\begin{align}
		\frac{g}{2} = \frac{1}{4\pi\varepsilon_{0}\varepsilon_{r}} \l( d_{1,\perp} d_{2,\perp} - 3 d_{1,\parallel} d_{2,\parallel} \r) \,,
	\end{align}
	we can estimate the maximum distance to effect a minimum coupling strength of $g_{\text{Min}} = 1$~MHz 
	\ifthenelse{\boolean{app}}{(c.f. Fig.~\ref{fig:TLSEnergy})}{(comparable to what is shown in Fig.~3 in the main text) }
	as $r_{\text{Max}}\approx 110\times 10^{-9}$~m.
	The volume in which TLS are interacting strongly enough is thus $V_{TLS} \sim 5.6\times 10^{-21}$~m$^{3}$.
	Assuming an overall TLS density of $10^{2}/(\mu\text{m}^{3}\text{GHz})$~\cite{Shalibo:PRL:2010, Barends:PRL:2013}, this leads
	to the effective frequency density of TLS in the interaction region of a single TLS of $\rho \sim 10^{-1} / \text{GHz}$.
	We note that the density obtained in Refs.~\onlinecite{Shalibo:PRL:2010, Barends:PRL:2013} refers only to high-frequency TS, 
	and a much higher density is expected for low-frequency TF~\cite{Simmonds:PRL:2004, Shnirman:PRL:2005}.

\ifthenelse{\boolean{app}}{\section*{Appendix D: Rabi-spectroscopy}}{\section{Rabi-spectroscopy}}

	When using non-frequency-tuneable qubits like single junction transmons, it is still possible to probe the form of the noise spectrum in close spectral vicinity of the qubit transition frequency. 
	To this end one can make use of the fact that for a driven system, 
	the frequencies of the noise spectrum relevant for decoherence will be shifted by the applied driving frequency.
	This effect can be thought of as a result of interaction of the dressing of the system states with drive photons, or similarly in the context of sideband transitions.
	The following derivation is based on the work in Ref.~\cite{Hauss:NJP:2008}, more details can be found there.
	
	For a two-level system driven with Rabi driving strength $\Omega_{0}$ at frequency $\omega_{d}$ we write the Hamiltonian
	\begin{align}
		\H = \frac{1}{2} \omega_{q} \sigma_{z} + \Omega_{0} \cos{\omega_{d} t} \: \sigma_{x} + \H_{\text{Sys-B}} + \H_{\text{B}} \,,
		\label{eq:HRabi}
	\end{align}
	with the qubit level-splitting $\omega_{q}$, the bare Rabi frequency $\Omega_{0}$ and driving frequency $\omega_{d}$.
	For the system-bath coupling term, we take
	\begin{align}
		\op H_{\text{Sys-B}} = \frac12 b_{\parallel} \sigma_{z} \op X_{\parallel} + \frac12 b_{\perp}\sigma_{x}\op X_{\perp} \,,
	\end{align}
	where the qubit level splitting is coupled to the bath variable $\op X_{\parallel}$ with coupling strength $b_{\parallel}$ and additionally the bath variable $\op X_{\perp}$
	might induce transitions between the qubit states due to its coupling with strength $b_{\perp}$. 
	Here the bath coupling constants $b$ are assumed to be small with respect to the other energies in the problem, 
	such that we can use perturbation theory in the strength of the system-bath coupling term $\H_{\text{Sys-B}}$.
	We will not specify the exact form of the bath Hamiltonian $\H_{\text{B}}$ but simply assume that is of a suitable form to induce Markovian decoherence, 
	i.e. it possesses a very large number of degrees of freedom and equilibrates on a time scale that is much shorter than all system time scales.
	Moving into a rotating frame at the drive frequency, we then find the decoherence rates as
	\begin{align}
		\Gamma_{\varphi} &= \sin^{2}{\beta}\: \gamma_{\varphi} + \frac12 \cos^{2}{\beta}\: \gamma_{1} \frac{ S_{X_{\perp}}(\omega_{d}) }{ S_{X_{\perp}}(\omega_{q}) } \,,\nn\\
		\Gamma_{\downarrow} &= \frac12 \cos^{2}{\beta}\: \gamma_{\Omega} 
			+ \frac14 \l( 1 - \sin\beta \r)^{2} \gamma_{1} \frac{ S_{X_{\perp}}(\omega_{d} + \Omega) }{ S_{X_{\perp}}(\omega_{q}) } \,,\nn\\
		\Gamma_{\uparrow} &= \frac12 \cos^{2}{\beta}\: \gamma_{\Omega} 
			+ \frac14 \l( 1 + \sin\beta \r)^{2} \gamma_{1} \frac{ S_{X_{\perp}}(\omega_{d} - \Omega) }{ S_{X_{\perp}}(\omega_{q}) } \,,
	\end{align}
	where we defined the rates
	\begin{align}
		\gamma_{\varphi} &= \frac12 b_{\parallel}^{2} S_{X_{\parallel}}(0) \quad \,,\quad 
		\gamma_{\Omega} = \frac12 b_{\parallel}^{2} S_{X_{\parallel}}(\Omega) \quad \,, \nn\\ 
		\gamma_{1} &= \frac12 b_{\perp}^{2} S_{X_{\perp}}(\omega_{q}) \,,
	\end{align}
	and we used the Rabi-frequency $\Omega = \sqrt{\Omega_{0}^{2} + (\omega_{q} - \omega_{d})^{2}}$. 
	Here, we introduced the symmetrized correlation functions for the bath variables $\op X$, defined as
	\begin{align}
		S_{X} (\omega) &= \frac12 \l( C_{X}(\omega) + C_{X}(-\omega) \r) \,.
	\end{align}
	where $C_{X} (\omega) = \int_{-\infty}^{\infty} dt \: \ee^{-\ii \omega \tau} \mean{X(\tau) X(0)}_{\text{th}}$ and the average $\mean{..}_{\text{th}}$ is over the steady state of the bath.
	For an environment in thermal equilibrium, the unsymmetrized noise spectrum will follow a detailed balance relation, 
	$C_{X} (\omega) = \ee^{-\beta\omega} C_{X}(\omega)$, with the inverse temperature $\beta = 1/k_{B}T$.
	The angle $\beta$ in these expressions defines the relationship between drive strength $\Omega_{0}$ and detuning between drive frequency and qubit splitting 
	and is defined as $\tan{\beta} = \Omega_{0} / (\omega_{q} - \omega_{d})$.
	
	The two rates $\gamma_{\varphi}$ and $\gamma_{1}$ can be determined in independent experiments, measuring relaxation from decay of the qubit excited state and decay of Ramsey fringes. 
	$\gamma_{\Omega}$ on the other hand can potentially be estimated using $\gamma_{\varphi}$ and assuming a $1/f$-type dependence of the low-frequency noise spectrum.
	
	In a Rabi experiment, the decay of the oscillations will be proportional to $\ee^{-\Gamma_{2} t}$ with $\Gamma_{2} = \Gamma_{\varphi} + \frac12 (\Gamma_{\uparrow} + \Gamma_{\downarrow})$ 
	and thus measurements of the Rabi oscillations at different drive strengths can be used to infer the noise spectrum in the vicinity of the qubit transition frequency $\omega_{q}$.

\ifthenelse{\boolean{app}}{\section*{Appendix E: Quasiparticle density fluctuations}}{\section{Quasiparticle density fluctuations}}

	Experimentally it was found that the temperature dependence of the relaxation rates of superconducting qubits could be well explained when assuming interacting with thermally excited quasiparticles~\cite{Catelani:PRB:2011}. 
	In this theory, the low temperature limit of the relaxation time $T_{1}$ stems from assuming a remaining density of non-equilibrium quasiparticles, the origin of which is not yet understood.
	Following the ideas developed in Ref.~\cite{Catelani:PRB:2011}, we conjecture that a fluctuating quasiparticle density, i.e. due to recombination events or tunnelling to an outside reservoir, 
	might lead to the observed fluctuations in relaxation time $T_{1}$. We calculate the required fractional changes in density as well as in terms of absolute number of quasiparticles for a
	given qubit design.
	
	The following calculations follow closely the theory of Ref.~\cite{Catelani:PRB:2011}, and we here only repeat their main steps for clarity.
	To derive the effects of the interaction between quasiparticles and superconducting circuits, 
	we start with a low-energy Hamiltonian describing tunnelling of quasiparticles across a Josephson junction at phase difference $\varphi$
	\begin{align}
		\H_{T} = \ii\: t \sum_{n,m,\sigma} \sin{\frac{\varphi}{2}} a^{L}_{n,\sigma}\!\hc a^{R}_{m,\sigma} + \text{h.c.}
		\label{eq:HTunnel}
	\end{align}
	where $t$ is the tunnelling amplitude and the operators $a^{L/R}_{n,\sigma}$ destroy a quasiparticle in state $n$ with spin $\sigma$ in the left/right lead.
	Eq.~\eqref{eq:HTunnel} is valid as long as the qubit energy $\omega$ as well as the characteristic energy $\delta E$ of the quasiparticles is much smaller than the superconducting gap $\Delta_{\text{sc}}$, 
	a condition which is well satisfied in experiments.
	Starting from this equation, the authors in Ref.~\cite{Catelani:PRB:2011} 
	derive the quasiparticle linear response function and thus the complex admittance of the Josephson junction due to quasiparticle tunnelling.
	
	Using the golden rule, we write the transition rates between qubit states due to quasiparticle tunnelling as
	\begin{align}
		\Gamma_{i\rightarrow f} = \abss{\bra i \sin{\frac{\varphi}{2}} \ket f} S_{qp}(\omega_{if})
	\end{align}
	where $\omega_{if} = \omega_{i} - \omega_{f}$ is the energy splitting between qubit states $\ket i$ and $\ket f$ 
	and $S_{qp}(\omega)$ is the quasiparticle spectral density, which can be calculated from the complex admittance via the fluctuation-dissipation theorem.
	For low temperature, $T \ll \Delta_{\text{sc}}$ and high frequencies $\omega_{if} \gg \delta E$, one finds
	\begin{align}
		S_{qp}(\omega) \approx x_{qp} \frac{8 E_{J}}{\pi} \sqrt{\frac{2 \Delta_{\text{sc}}}{\omega}}
	\end{align}
	with the junction's Josephson energy $E_{J}$ and the fractional quasiparticle density normalized to the density of Cooper pairs $x_{qp} = n_{qp} / 2 \nu_{0} \Delta_{\text{sc}}$.
	Here $\nu_{0}$ is the density of states of electrons in the leads, which we assume to be the same on both sides.
	
	For the case relevant to experiments, where a single junction 3D-transmon was used, the relaxation rate due to quasiparticles can then be calculated as
	\begin{align}
		\Gamma_{1\rightarrow0} = \frac{\omega_{p}^{2}}{\omega_{q}} \frac{x_{qp}}{2\pi} \sqrt{\frac{2\Delta_{\text{sc}}}{\omega_{10}}}
		\label{eq:Gamma1QP}
	\end{align}
	with the junction's plasma frequency $\omega_{p} = \sqrt{8 E_{J} E_{C}}$ and its charging energy $E_{C}$.
	Eq.~\eqref{eq:Gamma1QP} directly relates a qubit's relaxation rate to the density of quasiparticles.	
	From Eq.~\eqref{eq:Gamma1QP}, we can extract the fractional quasiparticle density $x_{qp}$, 
	with the value for the superconducting gap of thin-film aluminium~\cite{Paik:PRL:2011}:
	\begin{align}
		\Delta/2\pi &\approx 200~\mu\text{eV} \approx 50~\text{GHz} \approx 3.2 \times 10^{-23}~\text{J} \,.
	\end{align}
	Then, for a relaxation time of $T_{1} = 100 \mu$s, corresponding to $\Gamma_{1} = 10\times10^{3}/$s 
	\ifthenelse{\boolean{app}}{(c.f. Fig.~\ref{fig:Gamma1Time} in the main text)}{(c.f. Fig.~2 in the main text)}
	we find the canonical value of $x_{qp} \approx 5\times 10^{-7}$.
	
	We want to use the relative quasiparticle density determined above to calculate the actual number of quasiparticles interacting with the qubit sample. 
	For this we need the electron density of states at the Fermi edge for aluminum, which we take from literature as $\nu_{0} = 4.65 \times 10^{47} m^{-3} J^{-1}$~\cite{Court:PRB:2008}. 
	We thus find the quasiparticle volume density for the above used relaxation rate $\Gamma_{1} = 10\times10^{3}/$s as 
	\begin{align}
		n_{qp} = 2 \nu_{0} \Delta_{\text{sc}} x_{qp} \approx 5 \times 10^{18} \text{m}^{-3} \,.
	\end{align}
	 
	The 3D-transmon used in the experiments consists of two paddles of dimensions $350 \times 10^{-6} \cdot 700 \times 10^{-6} \cdot 120 \times 10^{-9} \text{m}^{3}$, 
	with a total volume of $V_{Al} \sim 3 \times 10^{-14} \text{m}^{3}$.
	We then find
	\begin{align}
		\delta N_{qp} = V_{Al} \: n_{qp} / 10 \approx 1.5 \times 10^{4}
	\end{align}
	as the number of quasiparticles that, for the sample used, leads to a change in the relaxation rate of $\delta\Gamma_{1} = 1\times10^{3}/$s. 

\ifthenelse{\boolean{app}}{}{	
	\bibliographystyle{apsrev4-1}
	\bibliography{T1Fluctuations}
}

\end{document}